\documentstyle[aps,amssym,12pt]{revtex}
\def\ha{Ha\-mil\-to\-nian}
\def\R{\Bbb R} 
 
\def\N{\Bbb N}
\def\T{\Bbb T}  

\def\la{\langle}
\def\be{\begin{equation}}
\def\ee{\end{equation}}
\def\ra{\rangle}
\def\ds{\displaystyle}
\begin{document}
\draft
\title{Deterministic spin models with a 
 glassy phase transition}

\author{
I.Borsari, M.Degli Esposti, S.Graffi and F.Unguendoli
 \\Dipartimento di Matematica, Universit\`{a} di Bologna
\\40127 Bologna, Italy}
\maketitle 
\vskip 12pt\noindent 
\begin{abstract}
 { We consider the infinite-range deterministic spin
models with
\ha\ $H=\sum_{i,j=1}^N J_{i,j}\sigma_i\sigma_j$, where $J$
is  the quantization of a chaotic map of
the torus.  The mean field  (TAP) equations
are derived by summing the high temperature expansion.
They predict a glassy phase transition at the critical
temperature  $T\sim 0.8$.}
\end{abstract}
\vskip 12pt\noindent

A class of infinite-range deterministic Ising spin models 
with glassy behaviour in numerical simulation has 
been recently
identified \cite{BM},\cite{MPR1},\cite{MPR2}: 
 unlike the random coupling case, however, 
the mean field equations \cite{PP}  
(hereafter the PP equations) are different from the
standard TAP ones\cite{TAP} and their linearization 
does not determine
a critical temperature for the glassy transition. \par 
The most interesting representatives of the class are the {\it
sine} (equivalently, {\it cosine}) models, with \ha  

\be 
\label{Ha}
H=-1/2\sum_{i,j=1}^N \,J_{i,j}\sigma_i\sigma_j
\ee
 
where the coupling matrix $J$ is (twice the uppermost left
block) of the  discrete sine (cosine) Fourier
transform  

\be
\label{J}
J_{i,j} =\frac{2}{\sqrt{2N+1}}\;
\sin\left(\frac{2\pi i j}{2N+1}\right), \quad 
i,j=1,\ldots,N 
\ee

In fact, the ground state of $H$ can be
explicitly computed\cite{MPR2} if $2N+1$ is prime
with $N$ odd: this leads to detect a
first-order crystalline phase transition (at a
higher critical temperature than 
the glassy one\cite{MPR1,MPR2}), whose relevance
on the glassy  behaviour of the system is discussed in
\cite{PP}, \S3. In turn, the matrix $J$
coincides\cite{BGU} with   (the imaginary part of)
the unitary propagator quantizing the discrete
dynamics generated by the unit symplectic matrix 
$S=\Bigg(\begin{array}{cc} 0 & -1 \\ 1 & 0
\end{array}\Bigg)$ acting as a \ha\ map over the
$2-$torus $\T^2$ (the operator quantizing a
\ha\ map of the torus  is a $N\times N$ unitary
matrix\cite{HB,DE}, $N$ being the inverse of the
Planck constant:\footnote{The physical intuition is:
the phase space has volume $1$, and can accommodate at
most $N$ quantum states of volume $\hbar$, so that 
$N\hbar=1$.} in this context the
thermodynamic limit $N\to\infty$ is formally
equivalent  to the classical limit). This algebraic
identity suggests as more natural candidates for detecting
deterministic glassy behaviour the coupling matrices
defined by the quantization of hyperbolic maps over
$\T^2$. The corresponding
 discrete dynamical systems are indeed chaotic, while
$S$ generates a period-$4$ one. Here we consider
the
 matrices 
$A=\Bigg(\begin{array}{cc} a & b \\ c & d
\end{array}\Bigg)$ with $a=d=2g, b=1, c=4g^2-1,
g\in\N$, which admit two positive eigenvalues
$\lambda_1>1, \lambda_2<1$.    The quantization of
 $A$ is  \cite{HB,DE,DGI}  the
unitary $N\times N$ matrix   
\be 
\label{V_A} 
V(A)_{jk}= {C_N}{N}^{-1/2}\exp {[({2\pi
i}/{N})(gj^2-jk+gk^2)]} 
\ee 
with 
$|C_N|=1$. The present models are thus defined by 
the  \ha s
$\ds H_A(\sigma)=\sum_{j,k}J(A)_{jk}\sigma_j\sigma_k$ 
with $J(A)={\rm Re}[V(A)]$, i.e.: 
\be 
\label{J_A}
 J(A)_{jk}={C_N}{N}^{-1/2}\cos
{[({2\pi i}/{N})(gj^2-jk+gk^2)]}
\ee
 Now 
$\overline{V}(A)=\overline{V}^T(A)={V}^{-1}(A)$
whence \footnote{This holds only if
$a,b,c,d$ are as above, and motivates our
 choice of hyperbolic matrices.}
  $\sigma({\rm Re}[ V(A)])={\rm Re}
[\sigma(V(A))]$.  Hence
the equistribution\cite{K} of the spectrum 
$\sigma(V(A))$ over the unit circle for $N\to\infty$
with an eigenvalue at $1$ for all $N$ implies the
same properties for
$\sigma(J(A))$  in
$[-1,1]$. \par Our claim is that these deterministic
models behave, at least in mean field
approximation, more closely to the random ones, such
as the SK one\cite{SK}. We will indeed compute in
closed form (at the thermodynamic limit
$N\to\infty$) the Gibbs
(i.e., magnetization dependent) free energy
$\beta\Phi$. The result is:    
$$
\beta\Phi = 1/2\sum_{i=1}^N\left\{(1+m_i){\rm
ln}(1+m_i)/2+ (1-m_i){\rm
ln}(1-m_i)/2\right\}-
$$
\be
 \label{FE}
-\beta/2\sum_{i,j=1}^NJ(A)_{ij}m_im_j-
NG(\beta(1-q)); \;G(\beta)={\beta^2}/({8+4\beta^2})
\ee
where $m_l:l=1,\ldots,N$ is the magnetization at site
$l$ and $\ds q=1/N\sum_{l=1}^nm_l^2$ the
Edwards-Anderson order parameter (as
in \cite{PP}, $G(\beta)$  does not depend on the
particular choice of $J(A)$). The stationarity condition
of the Gibbs free energy\cite{MPV} yields the mean field
 equations of the model (believed exact because of its
infinite range) 
\be
\label{TAP}
Q_i\equiv \tanh^{-1}m_i+2\beta
G^{\prime}(\beta(1-q))m_i 
    -\beta\sum_j J(A)_{ij}m_j=0
\ee
Unlike the PP ones (but like the TAP\cite{TAP}) 
these equations can be solved  by linearization 
near $q=0$: any  eigenvector for  the
 eigenvalue $1$ of $J(A)$ yields a
solution if $\ds {2\beta^2}/{(2+\beta^2)^2} -\beta+1
=0$.
%
%
%
%
Its only positive zero $\beta \sim 1.25$ determines
the critical temperature $T\sim 0.8$. The phase
transition is glassy because, as we will see, $\ds
1/N\sum_{l=1}^Nm_l\to 0$ as $N\to\infty$. This fact
and (\ref{FE}) implies that all such solutions have
the same free energy. The thermodynamics is
thus independent of their number, the
multiplicity of
 the eigenvalue $1$, which depends in a 
sensitive way on the integer
$N$\cite{K}.\par To get the mean field equations
(\ref{TAP}) we  resum the high temperature
expansion by the same procedure of \cite{PP}. The
function
$G(\beta)$ has a simpler form in this case because
there is only one class of non vanishing
diagrams at the thermodynamic limit.\par\noindent 
Consider the Helmholtz free energy
$F(\beta)$: if $ Z(\beta)=\sum_{\{\sigma\}}\exp{\beta
H_A(\sigma)/2}=\exp{-\beta F(\beta)}$ is the
 partition
function (at site-dependent magnetic field $h_i=0$)
we have\cite{ID} 
$$
e^{-\beta F}= {(2/\pi)^{N/2}}/
{\rm
det}^{\frac12}(\beta J)\int_{\R^N} \exp\{
  \la (2\beta J)^{-1} \phi, \phi\ra +
    \sum_{i}\log\cosh(\phi_i+h_i)\}\,{d^N\phi}|_{h_i=0}= 
$$
\be
 \label{HEF}  {2^N}{
\pi}^{-N/2} \int_{\R^N} \exp{\Big\{-<x,x> + \sum_{i=1}^N
\log\cosh\Big[\sqrt{2\beta}
  \sum_{h}((J^{1/2})_{ih}x_h\Big]\Big\}}\,{d^Nx}
\end{equation}
The high temperature expansion for $\beta F(\beta)$ is
generated out of the integration of the expansion of 
$\exp{\log\cosh x}$ . The well known
diagrammatic representation of its $n$-th order term
is obtained\cite{ID} by drawing all diagrams with $n$
links, $2\leq j+1\leq n+1$ vertices and no
external legs, whose  individual contribution is
 \par\noindent
1. 
For any link between two consecutive vertices $l\neq k$ 
a factor $\beta J_{lk}$;
\par\noindent 2. For any vertex with $m$ links the cumulant
$u_m$, i.e. the $m-$th coefficient of the Taylor expansion
of $\log\cosh x$ (\cite{ID}, p.414); 
\par\noindent 3. Any diagram has to be divided by its
order of symmetry.
\par
 The contribution of each individual
diagram $D$ at order $\beta^n$ is indeed 
\be
\label{diagram}
|D|=U(D)S(D)^{-1}\sum_{r_1,\ldots,r_{j+1}}
J_{r_1,r_2}^{\alpha_1}\cdot
J_{r_2,r_3}^{\alpha_2}\cdots
J_{r_j,r_{j+1}}^{\alpha_j}, \qquad j=1,\ldots,n \ee
Here $j+1$ is the number of
vertices, $\alpha_1+\ldots+\alpha_j=n$; 
$n-j+1=n-1,\ldots,1$ is  the number of loops,  $\alpha_i
\geq 1$ the number of links between consecutive vertices,
 $S(D)$ the  symmetry factor, and $U(D)= 
{u_1(\alpha_1)u_2(\alpha_1+\alpha_2)\cdots
u_j(\alpha_{j-1}+\alpha_j)u_{j+1}(\alpha_j)}$. Remark that
$r_{n+1}=r_1$ for $j=n$. 
We now verify that,
at the thermodynamic limit $N\to\infty$:\par\noindent
(a) All diagrams
 for $n=2p+1$ vanish;
\par\noindent
(b) For $n=2p$ the only surviving
diagram $D_p$ is the one with $p+1$ vertices, 
 $p$ loops and two links between consecutive loops:
 \par\noindent
\begin{picture}(800,30)
\put(20,10){\circle*{5}}
\put(30,10){\circle{20}}
\put(40,10){\circle*{5}}
\put(50,10){\circle{20}}
\put(60,10){\circle*{5}}
\put(60,10){.....}
\put(80,10){\circle*{5}}
\put(90,10){\circle{20}}
\put(100,10){\circle*{5}}
\put(120,5){$:\vert D_p\vert= (-1)^{p-1}2^{p-2}$}

\end{picture}
To
prove (a) and (b), first recall the basic
estimate fulfilled by the Gauss sums\cite{Ap}
\be
\label{Gauss}
\Big|\sum_{s_1,\ldots,s_l=1}^N\exp{({2\pi i}/{N})
g(s_1,\ldots,s_l)}\Big|\leq C N^{{l}/{2}}
\ee
where $g$ is any quadratic form in the $l$ integers
$s_1,\ldots,s_l$ with integer coefficients and $C$ a
constant independent of
$g$ and
$N$. By (\ref{V_A}) and (\ref{J_A}), the
sum in (\ref{diagram}) amounts for any fixed $1\leq j \leq
n$ to $2^n$ Gaussian sums over $r_1,\ldots,r_j$ divided by
$N^{n/2}$.  We now argue that all these
 Gaussian sums are, once divided by $N^{n/2}$, of
order $1$ or less except one: that with $n=2p$,
$\alpha_1=\ldots=\alpha_p=2$, and summand
independent of $r_1,\ldots,r_{p+1}$, generated by the
constant term in the expansion of $\ds \prod_k
J^2_{r_k,r_{k+1}}$.  This sum is clearly  equal to
$\ds N^{p+1}N^{-p}2^{-p}= {N}2^{-p}$.  Any other
(divided) sum is down by at least ${N}^{-1}$: each
summation over $q$ indices is estimated by $N^{q/2}$
if the summand is a Gaussian and by $N^q$ if the
summand is constant, which case shows up whenever
$\alpha_i$ is even for at least one $i$; on the other
hand any power of $\alpha_i$ in excess of one
increases by one the number of loops and thus reduces
by one the number of vertices and hence of summation
indices. Thus for $n=2p+1$ the most divergent sums
behave  as $N^{p+1/2}$: this happens either for
$\alpha_i=1$ for all $i$ so that $j=2p$, or for
$\alpha_1=\ldots=\alpha_{p-1}=2, \alpha_{p}=3$ (and
permutations thereof) so that $j=p+1$. The prefactor
$N^{-(p+1/2)}$ yields the estimate $O(1)$. For $n=2p$
the second most divergent sums have the same
behaviour: again they take place for $\alpha_i=1$
for all $i$ and $j=2p-1$ or for  $\alpha_1=\ldots=
\alpha_{p}=2$ with the summand depending but on two
indices (up to permutations).  Now 
the only surviving $D_p$ has two vertices with two
links (the extrema) and $p-1$ vertices with four
links (all the remaining ones). Since $u_2=1,
u_4=-2$, $U(D_p)=(-2)^{p-1}$. Moreover the symmetry
factor $S(D_p)$ is $2\cdot 2^p$ (to account for the
interchange of the external vertices and of any pair of
links between the internal vertices). Hence
$U(D_p)/S(D_p)=(-1)^{p-1}/4$ and by (\ref{diagram}) we
have  $|D_p|=(-1)^{p-1}\cdot 2^{-p-2}$. Summing up we get
the Helmholtz free energy:  
\be
\label{somma}
-\beta F(\beta)-N{\rm ln}2=N\sum_{p=1}^{\infty}
(-1)^{p-1}\cdot
2^{-p-2}\beta^{2p}=N{\beta^2}/({8+4\beta^2})=NG(\beta)
\ee
As in \cite{PP}, expanding $-\beta F(\beta)$ to first
order we recover the SK Helmholtz free energy $N{\rm
ln}2+\beta^2/4\sum_{i,j}J_{i,j}^2$ because here 
$\sum_{i,j}J_{i,j}^2= N/2+O(1)$ for $N$ large.
\par
To obtain the Gibbs free energy we have to  
perform the Legendre transform    
\be
\label{LT}
\Phi(\beta,m_i)\equiv
{\rm max}_{h_i}[F(\beta,h_i)-\sum_ih_im_i]
\ee
out of the ($h_i$ dependent) expansion of (\ref{HEF}) in
powers of
$\beta$.  To this end we simply take over the
 Parisi-Potters argument because we are
summing over a subclass of the "cactus" diagrams
considered in
\cite{PP} within the same assumption of self-averaging,
namely $\ds m_k^2= N^{-1}\sum_{l=1}^Nm^2_l +o(1)$ as
$N\to\infty$. This yields (\ref{FE}), whence the mean
field equations (\ref{TAP}) and the critical
temperature.\par\noindent It remains to prove that
$\ds N^{-1}\sum_{l=0}^{N-1}m_k^l\to 0$ as
$N\to\infty$  if $J(A)m_k=m_k$.
 Any such eigenvector defines indeed a critical
(staggered) magnetization distribution; furthermore the
vanishing of the total magnetization represents the
necessary condition both for the glassy nature of the
transition as well as for self-averaging. Setting
$\phi_0=N^{-1/2}(1,1\ldots,1)$ we have  $\ds
N^{-1}\sum_{l=0}^{N-1}m_k^l=\la m_k,\phi_0\ra= \la
m_k,V(A)^l\phi_0\ra\;\forall l\in\N$. On the other
hand\cite{K} $V(A)$ tends (weakly) to a unitary
operator on $L^2(-\pi,\pi)$ with Lebesgue spectrum on
the unit circle as $N\to\infty$. This
entails\cite{CFS}  $\ds
\lim_{l\to\infty}\lim_{N\to\infty}\la
m_k,V(A)^l\phi_0\ra=\lim_{N\to\infty}\la
m_k,\phi_0\ra=0$. \par\noindent Moreover by
(\ref{FE}) the specific free energy $f=\beta\Phi/N$
is the same for all eigenvectors $m_k$: the
third term depends only on $q$, the second is $-\beta
q/2$ because $J m_k=m_k$ and  
$$
{2N}^{-1}\sum_{i=1}^N\left\{(1+m_i){\rm
ln}(1+m_i)/2+ (1-m_i){\rm
ln}(1-m_i)/2\right\}=q+O(q^{3/2})
$$
since $q$ is by definition small near the critical
point. 
We conclude with two remarks.\par\noindent
(i) The staggered magnetizations can be explicitly
computed for some particular values of $N$. There is
indeed\cite{K} 
$p(N)\in\N$ such that
$J(A)^p=I_d$, and for "most" sequences of values of $N$
one has
$p(N)/N=M, M<\infty$. Under these conditions the
eigenvalue
$1$ of
$J(A)$ has multiplicity $M+1$. A first corresponding
(norm
$\epsilon$) eigenvector is\cite{DGI} $m_1=\cos{({2\pi
i}/{N})
\overline{k}l^2}/\sqrt{\epsilon N}$ where
$\overline{k}^2=3\;({\rm mod} N)$, while 
$$
m^r_l=1/\sqrt{\epsilon 
p}\sum_{s=0}^{p-1}c_s^r \cos{({2\pi
i}/{N})(a_s^rl^2+b_s^rl})/\sqrt{N},\qquad
r=2,\ldots,M+1
$$
where $|c_s^r|\leq 1$ and $0\leq a_s^r,b_s^r\leq N-1$ are
integers (\cite{DGI}, Formula (4.25)). By
 (\ref{Gauss}) one checks
directly that $\ds N^{-1}\sum_{l=0}^{N-1}m_l^r\to 0$ for
all eigenvectors.
\par\noindent (ii)
The magnetization just below the critical
point has the square-root behaviour of a
second order transition.  Compute indeed the 
second order
expansion $\delta^2Q_i$ near $\beta_c$ and $m^*$
 as in  \cite{TAP}, putting
$\beta=\beta_c+\Delta\beta$ and 
$m_i=m^*_i+\delta m_i$ where $m^*$ is any
staggered magnetization vector and $\la\delta
m,m^*\ra=0$. Neglecting the term of order
$\Delta\beta^2$, taking the
scalar product with $m^*$ and dividing by $Nq$
the equation $\la m^*,\delta^2Q\ra=0$ we get $\ds
\alpha\|\delta m\|^2+\gamma \Delta\beta=0$
with $\ds
\alpha=-2\beta_c^3G"(\beta_c)>0,
\gamma=-1+2G'(\beta_c)+2\beta_cG"(\beta_c)<0$,
whence the assertion.
\vskip 0.2cm\noindent
{\bf Acknowledgment.} We thank G.Parisi
for a useful conversation.



\begin{thebibliography}{DEGL}   
 \bibitem[1]{BM} {\sc
J.-P.Bouchaud and M.M\'ezard},   
J.Phys.I (France) {\bf 4} (1994), 1109 
\bibitem[2]{MPR1} {\sc
E.Marinari, G.Parisi and F.Ritort}, 
J.Phys.A (Math.Gen.) {\bf 27} (1994),
7615  
\bibitem[3]{MPR2} {\sc E.Marinari, G.Parisi and
F.Ritort}, 
J.Phys.A (Math.Gen.) {\bf 27} (1994), 7647 
\bibitem[4]{PP} {\sc G.Parisi and M.Potters}, 
J.Phys.A (Math.Gen.) {\bf 28} (1995), 5267 
\bibitem[5]{TAP} {\sc J.Thouless, P.W.Anderson and
T.Palmer},  Phil.Mag. {\bf 35} (1977), 593 
\bibitem[6]{BGU} {\sc I.Borsari, S.Graffi and
F.Unguendoli},  J.Phys.A (Math.Gen.), to appear
\bibitem[7]{HB} {\sc J.Hannay and M.V.Berry}, 
 Physica D {\bf  1} (1980), 267
\bibitem[8]{DE} {\sc M.Degli Esposti}, 
Ann.Inst.H.Poincar\'e {\bf 58} (1993), 7647  
\bibitem[9]{DGI} {\sc M.Degli Esposti, S.Graffi
and S.Isola}, 
Commun.Math.Phys. {\bf 167} (1995), 471
\bibitem[10]{K} {\sc J.Keating} Nonlinearity  {\bf  4} ( 
1991), 309
\bibitem[11]{SK}\ {\sc D.R.Sherrington and
A.L.Kirckpatrick}, Phys.Rev.Letters  {\bf 35}, (1975),
1792 
\bibitem[12]{MPV} {\sc M.M\'ezard, G.Parisi and
M.Virasoro},  \ {\it Spin Glass Theory and Beyond},
World Scientific, Singapore (1985), 
\bibitem[13]{ID} {\sc C.Itzykson, J.M.Drouffe}, {\it
Statistical Field Theory}, (Vol.II), Cambridge University
Press, 1988 
 \bibitem[14]{Ap} {\sc T.Apostol}, \
{\it Introduction to Analytic Number Theory},
Springer-Verlag, New York 1976 
 \bibitem[15]{CFS} see e.g. {\sc I.P.Cornfeld,
S.V.Fomin and Ya.G.Sinai}, \ {\it Ergodic Theory},
Springer-Verlag, New York 1986 

\end{thebibliography}
\end{document}